# ANALYZING THE PERFORMANCE OF PROBABILISTIC ALGORITHM IN NOISY MANETS

Hussein Al-Bahadili and Khalid Kaabneh

The Arab Academy for Banking & Financial Sciences, P.O. Box 13190, Amman 11942, Jordan
hbahadili@aabfs.org, kkaabneh@aabfs.org

## ABSTRACT

*Probabilistic broadcast has been widely used as a flooding optimization mechanism to alleviate the effect of broadcast storm problem (BSP) in mobile ad hoc networks (MANETs). Many research studies have been carried-out to develop and evaluate the performance of this mechanism in an error-free (noiseless) environment. In reality, wireless communication channels in MANETs are an error-prone and suffer from high packet-loss due to presence of noise, i.e., noisy environment. In this paper, we propose a simulation model that can be used to evaluate the performance of probabilistic broadcast for flooding in noisy environment. In the proposed model, the noise-level is represented by a generic name, probability of reception ($p_c$) ($0 \leq p_c \leq 1$), where $p_c=1$ for noiseless and $<1$ for noisy environment. The effect of noise is determined randomly by generating a random number $\xi$ ($0 \leq \xi < 1$); if $\xi \leq p_c$ means the packet is successfully delivered to the receiving node, otherwise, unsuccessful delivery occurs. The proposed model is implemented on a MANET simulator, namely, MANSim. The effect of noise on the performance of probabilistic algorithm was investigated in four scenarios. The main conclusions of these scenarios are: the performance of probabilistic algorithm suffers in presence of noise. However, this suffering is less in high density networks, or if the nodes characterized by high retransmission probability or large radio transmission range. The nodes' speed has no or insignificant effect on the performance.*

## 1. INTRODUCTION

A MANET is defined as a collection of low-power wireless mobile nodes forming a temporary network without the aid of any established infrastructure or centralized administration [1]. A data packet in a MANET is forwarded to other mobile nodes within the network through a reliable and an efficient route established by routing protocols [2]. The most widely used routing protocols in MANETs are known as dynamic routing protocols, such as ad hoc on-demand distance vector routing (AODV) [3], dynamic source routing (DSR) [4], and location-aided routing (LAR) [5]. Dynamic routing protocols consist of two major phases, route discovery and route maintenance [6].

Route discovery is the phase in which a route between source and destination nodes is established for the first time. In this phase, the source broadcasts a route request (RREQ) packet to its neighbors, which then forward it to their neighbors, and so on, until either the destination itself or a node that have a fresh route to the destination is located, which subsequently responds with a route reply (RREP) packet back to the source through the route from which it first received the RREQ. In the route maintenance phase, the route is maintained; and if it is broken for any reason, then the source either finds other known route on its routing table or initiates new route discovery procedure, therefore, the cost of information exchange during route discovery is higher than the cost of point-to-point data forwarding after the route is established [7].

Broadcasting is a fundamental communication primitive for route discovery in routing protocols in MANETs. One of the earliest broadcast mechanisms proposed in the literature is pure flooding, which is also called simple or blind flooding. Although it is simple and reliable, pure flooding is costly where it costs *n* transmissions in a network of *n* reachable nodes. In addition, pure flooding results in serious redundancy, contention, and collisions in the network; such a scenario has often been referred to as the broadcast storm problem (BSP) [8]. To eliminate the effects of BSP during route discovery in MANETs, a variety of broadcast optimization algorithms have been developed to reduce the number of retransmission during the route discovery phase, such as: probabilistic [1, 2], LAR [5, 8], multipoint relaying [9, 10], cluster-based [8, 11] algorithms. As the number of retransmissions is decreased, the bandwidth is saved and contention and node power consumption are reduced, and this will improve the overall network performance.

In this work, we concern with the probabilistic broadcast algorithm, which we shall refer to as probabilistic algorithm. In this algorithm, when receiving a RREQ packet, a node retransmits the packet with a certain retransmission probability ($p_t$) and with probability ($1-p_t$) it discards the packet. A node is allowed to retransmit a given RREQ packet only once, which identifies it through its sequence number.

Many research studies have been carried-out to develop and evaluate the performance of probabilistic algorithm in a noiseless environment, while in reality wireless communication channels in MANETs are unreliable and suffer from high packet-loss due to presence of noise. In this paper, we propose a simulation model that can be used to investigate the effect of noise on the performance of probabilistic algorithm. The proposed model is implemented on a MANET simulator (MANSim) [12].

In order to evaluate the performance of probabilistic algorithm, four scenarios were simulated. These scenarios investigated the variation of two computed parameters, namely, the number of retransmissions (RET) and the network reachability (RCH) against $p_c$ for various $p_t$'s, number of nodes (*n*), nodes speed (*u*), and nodes radio transmission range (*R*). The simulation results demonstrated that RCH decreases with increasing noise-level or decreasing $p_c$. Due to the fact that presence of noise increases packet-loss rate in the network or in other words it decreases $p_c$ of route request packets by neighboring nodes and subsequently the destination node, i.e., yields lower RCH. Thus, the RREQ packet needs to be re-initiated more than once before a route is established.

The rest of the paper is organized as the follows. The most recent and related work is presented in Section 2. In Sections 3 and 4, descriptions of pure and probabilistic algorithm in noiseless MANETs, are given. The wireless network environment and the effect of noise are described in Section 5. The description of the probabilistic algorithm in noisy MANETs is given in Section 6. Section 7 presents the description, results, and discussion of the four scenarios that have been performed to investigate the performance of the proposed algorithm. Finally, in Section 8, a number of conclusions are drawn and recommendations for future work are pointed-out.

## 2. RELATED WORKS

There are a number of mechanisms that have been developed to alleviate the effects of the BSP in MANETs by inhibiting some nodes from rebroadcasting to reduce number of retransmissions, and thus, collisions and contention [8, 13]. Since in this paper, we concern with probabilistic algorithm, in this section, we present a review of some of the most recent and related work on probabilistic algorithm in both noiseless and noisy MANETs. Probabilistic algorithm was used for ad hoc route discovery by Haas et. al [14], and it was called the gossip-based ad hoc route discovery (GOSSIP1) approach. GOSSIP1 has a slight problem with initial conditions. If the source has relatively few neighbors, there is a chance that none of them will gossip, and the gossip will die. Similar conclusions were also explored by Ni et. al [15]. To make sure this does not happen, Haas et. al proposed a modified protocol, in which they gossip with $p_t$=1 for the first *h* hops before continuing to gossip with $p_t$<1. Their results showed that they can save up to 35% message overhead compared to pure flooding. Furthermore, adding gossiping to a protocol such AODV and ZRP not only gives improvements in the number of messages sent, but also resulted in improved network performance in terms of end-to-end latency and throughput.

S. Tseng et. al [8] investigated the performance of the probabilistic algorithm for various network densities in noise-free environment. They presented results for three network parameters, namely, reachability, saved rebroadcast, and average latency, as a function of $p_t$ and network density. J. Kim et. al [16] introduced a dynamic probabilistic algorithm with coverage area and neighbors confirmation for MANETs. Their scheme combines probabilistic approach with the area-based approach. A mobile host can dynamically adjust $p_t$ according to its additional coverage in its neighborhood. The additional coverage is estimated by the distance from the sender. The simulation results showed this approach generates fewer rebroadcasts than pure flooding approach. It also incurs lower broadcast collision without sacrificing high reachability.

D. Scott an A. Yasinsac [2] presented a dynamic probabilistic solution that is appropriate to solving BSP in dense mobile networks, also referred to as gossip protocol. The approach can prevent broadcast storms during flooding in dense networks and can enhance comprehensive delivery in sparse networks. M. Bani-Yassein et. al [1, 17, 18, 19] proposed a dynamic probabilistic algorithm in MANETs to improve network reachability and saved rebroadcast. The algorithm determines $p_t$ by considering the network density and node movement. This is done based on locally available information and without requiring any assistance of distance measurements or exact location determination devices. The algorithm controls the frequency of rebroadcasts and thus might save network resources without affecting delivery ratios.

K. Viswanath, and K. Obraczka [20] developed an analytical model to study the performance of plain (pure) and probabilistic algorithms in terms of its reliability and reachability in delivering packets. They provided simulation results to validate the model. Their simulation results indicated that probabilistic algorithm can provide similar reliability and reachability guarantees as plain flooding at a lower overhead. Q. Zhang and D. Agrawal [21] proposed a probabilistic approach that dynamically adjusts $p_t$ as per the node distribution and node movement. The approach combines between probabilistic and counter-based approaches. They evaluated the performance of their approach by comparing it with the AODV protocol (which is based on pure flooding) as well as a fixed probabilistic approach. Simulation results showed that the approach performs better than both pure and fixed probabilistic algorithms.

J. Abdulai et. al [22] analyzed the performance of AODV protocol over a range of possible $p_t$. Their studies focused on the route discovery part of the routing algorithm, they modified the AODV routing protocol implementation to incorporate $p_t$. Results obtained showed that setting efficient $p_t$ for AODV routing discovery has a significant effect on the general performance of the protocol. The results also revealed that the optimal $p_t$ for efficient performance of the routing protocol is affected by the prevailing network conditions such as traffic load, node density, and node mobility. During their study they observed that the optimal $p_t$ is around 0.5 in the presence of dense network conditions and around 0.6 for sparse network conditions.

A. Hanashi et. al [23] proposed a dynamic probabilistic broadcast approach that can efficiently reduce broadcast redundancy in MANETs. The algorithm dynamically calculates $p_t$ according to the number of node's first-hop neighbors (*k*). The $p_t$ would be low when *k* is high, which means host is in dense area, and $p_t$ would be high when $k$ is low, which means host is in sparse area. They compared their approach against pure flooding approach, fixed probabilistic approach, and adjusted probabilistic flooding by implementing them in a modified version of the AODV protocol. The simulation results showed that broadcast redundancy can be significantly reduced through their approach while keeping the reachability high. It also demonstrated lower broadcast latency than existing approaches. I. Khan et. al [24] proposed a coverage-based dynamically adjusted probabilistic forwarding scheme and compared its performance with pure and fixed probabilistic schemes. The scheme keeps up the reachability of pure flooding while maintaining the simplicity of probability based schemes.

C. Barret et. al [25] introduced a probabilistic routing protocols for sensor networks, in which an intermediate sensor decides to forward a message with $p_t$ that depends on various parameters, such as the distance of the sensor to the destination, the distance of the source sensor to the destination, or the number of hops a packet has already traveled. They proposed two protocol variants of this family and compared the new methods to other probabilistic and deterministic protocols, namely constant-probability gossiping, pure flooding, random wandering, shortest path routing, and a load-spreading shortest-path protocol. The results showed that the multi-path protocols are less sensitive to misinformation, and suggest that in the presence of noisy data, a limited flooding strategy will actually perform better and use fewer resources than an attempted single-path routing strategy, with the parametric probabilistic sensor network routing protocols outperforming other protocols. The results also suggested that protocols using network information perform better than protocols that do not, even in the presence of strong noise.

## 3. PURE FLOODING

The simplest solution that can be used for broadcasting is known as pure flooding; blind flooding, or simple flooding, in which each node retransmits the received message when it receives it for the first time starting at the source node. This process continues until all reachable nodes have received and retransmit the broadcast message. Figure 1 outlines this pure flooding algorithm [1]. The major advantages of pure flooding are its simplicity and reliability. On the other hand, the main disadvantage is that it costs $n$ transmissions in a network of $n$ reachable nodes. Other main drawbacks of flooding in MANETs include [8, 13]:

(1) Redundant retransmissions: When a mobile node decides to retransmit a broadcast message to its neighbors, all its neighbors already have the message.
(2) Contention: After a mobile node retransmits a message, if many of its neighbors decide to retransmit the message, these transmissions (which are all from nearby nodes) may severely contend with each other.
(3) Collision: Because of the deficiency of back-off mechanism, the lack of request to send/clear to send (RTS/CTS) dialogue, and the absence of collision detection, collisions are more likely to occur and cause more damage.

The only requirement made for the retransmitting nodes in MANETs is that they assess a clear channel before retransmitting. Clear channel assessment does not prevent collisions from hidden nodes. Additionally, no alternative is provided for collision prevention when two neighbors assess a clear channel and transmit simultaneously. Due to the lack of RTS/CTS dialogue, a node has no way of knowing whether a massage was successfully delivered to its neighbors. In high density networks, a significant amount of collisions occur leading to many dropped packets. Therefore, effective and efficient flooding protocols, such as probabilistic flooding, are so important since it always try to limit the probability of collisions by limiting the number of retransmissions in the network.

```
On receiving a route request at node i, do the following:
    If IRet(i) = 0 Then {The node has not retransmitted the request before (IRet(i) = 0}
        Retransmit request
        IRet(i) = 1 {Update the node retransmission index IRet(i) by equating it to 1}
    End if
```

Figure 1. Pure flooding algorithm.

## 4. PROBABILISTIC ALGORITHM

Probabilistic algorithm is widely-used for flooding optimization during route discovery in MANETs [22-24]. It aims at reducing number of retransmissions, in an attempt to alleviate the BSP in MANETs [8]. In this algorithm, when receiving a RREQ packet, a node retransmits the packet with a certain $p_t$ and with probability $(1-p_t)$ it discards the packet. A node is allowed to retransmit a given RREQ packet only once, i.e., if a node receives a packet, it checks to see if it has retransmitted it before, if so then it just discards it, otherwise it performs its probabilistic retransmission check. Nodes usually can identify the RREQ packet through its sequence number. In the probabilistic algorithm, the source node $p_t$ is always set to 1, to enable the source node to initialize the RREQ. While, $p_t$ for intermediate nodes (all nodes except the source and the destination nodes) is determined using a static or dynamic approach. In this work, we concern with fixed retransmission probability. Figure 2 outlines the probabilistic algorithm.

```
On receiving a route request at node i, do the following:
    If IRet(i) = 0 Then {The node has not retransmit the request before (IRet(i) = 0}
        ξ = rnd() {ξ some random number between 0 and 1}
        If ξ <= p_t Then
            Retransmit request
            IRet(i) = 1 {Update the node retransmission index IRet(i) by equating it to 1}
        End if
    End if
```

Figure 2. Probabilistic algorithm

## 5. WIRELESS NETWORK ENVIRONMENT

The wireless network environment can be categorized, according to the presence of noise, into two types of environments; these are [9]:

(1) A noiseless (error-free) environment, which represents an ideal network environment, in which all data transmitted by a source is assumed to be successfully and correctly received by a destination node. It is characterized by the following axioms: the world is flat, all radios have equal range, and their transmission range is circular, communication link symmetry, perfect link, signal strength is a simple function of distance.
(2) A noisy (error-prone) environment, which represents a realistic network environment, in which the received signal may differ from the transmitted signal, due to various transmission impairments, such as: signal attenuation, free space loss, thermal noise, atmospheric absorption, multipath effect, refraction. In addition, error in reception may occur due to rapidly changing topologies that are caused by nodes movement. All of these impairments are represented by a generic name, noise, and the environment is called noisy environment. The noise-level in the noisy environment expressed in terms of a probability of reception ($p_c$), which is defined as the probability that a transmitted data is survived being lost and successfully delivered to its destination despite the presence of all/any of the above impairments.

## 5.1    Effect of Noise

This section shows how a noise negatively affects the performance of probabilistic algorithms. This can be explained with the help of Figure 3 as follows: Assume that nodes A and G are the source and destination nodes, and nodes B and C are intermediate nodes. Node G can be reached through nodes B and C. Figure 3a shows the RREQ packet dissemination from node A to G using pure flooding in noiseless environment, in which node G receives the RREQ packet twice through nodes B and C. However, it will send only one RREP through node B or C depending on which one forward the RREQ first.

On the other hand for the same network topology, if probabilistic flooding is used, there are three possibilities, these are: (i) both B and C, (ii) either B or C, and (iii) neither B nor C will retransmit the RREQ packet. Assume that only node B succeeds to retransmit the RREQ packet. For the same topology in noisy environment, assume that node A be unsuccessful in delivering the RREQ packet to node B (due to presence of noise) and be successful in delivering the RREQ packet to node C. Thus, the RREQ packet will not be delivered to node G, and node G appears as unreachable. This is because node B has no packet to retransmit and node C prohibits from retransmission by the probabilistic algorithm.

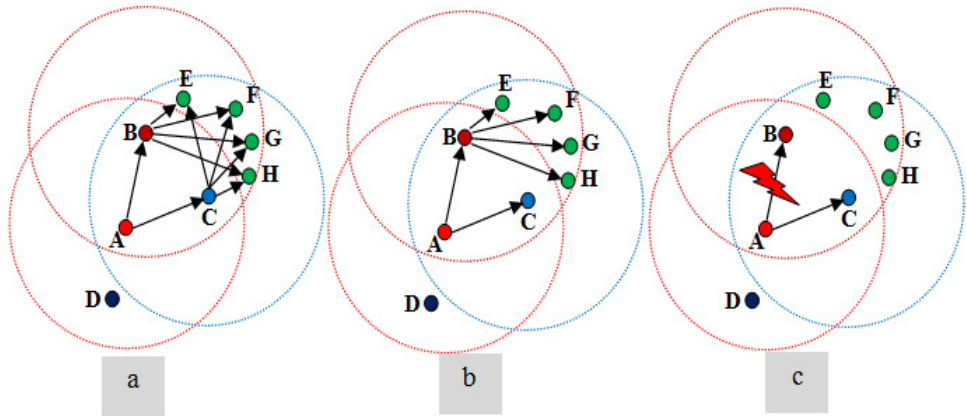

Figure 3. Pure and probabilistic algorithms in various network conditions: (a) Pure flooding algorithm, (b) Probabilistic algorithm in noiseless environment, and (c) Probabilistic algorithm in noisy environment.

## 6.   THE PROPOSED PROBABILISTIC ALGORITHM IN NOISY ENVIRONMENT

The probabilistic route discovery algorithm in noisy MANETs can be implemented as follows: When the distance between the transmitting and receiving nodes is less than the radio transmission range of the transmitting node, a random test is performed to decide whether the RREQ is successfully delivered to the receiving node or being lost due to error. The random test is performed by generating a random number $\xi$ ($0\leq\xi<1$) and compared it with $p_c$, if $\xi\leq p_c$, then the RREQ is successfully delivered to the receiving node; otherwise, the RREQ is not delivered or being lost.

The value of $p_c$ is either predetermined or instantly computed using a certain probability distribution function (PDF). If the RREQ packet is successfully delivered to the node, then the receiving node performs its probabilistic algorithm to find out whether it should retransmit the RREQ packet or not. Figure 4 outlines the proposed probabilistic algorithm in noisy MANETs.

```
Probabilistic Algorithm in Noisy Environment
   If (IRange = 1) Then {The receiving node is within the transmission range of the sender, in a noiseless
                         environment this guarantees request reception by the receiver, while in a noisy
                         environment a random test must be performed to find-out whether a successful
                         delivery occurs or not. IRange=0 means the receiver is not within the
                         transmission range of the sender}
      ξ₁ = rnd()  {ξ₁ some random number between 0 and 1}
      If (ξ₁ <= p_c) Then {Reception random test}
         IRec(i)++ {Update the node reception index IRec(i)}
         If (IRet(i) = 0) Then {The node has not retransmitted the request before (IRet(i) = 0)}
            ξ₂ = rnd() { ξ₂ some random number between 0 and 1}
            If (ξ₂ <= p_t) Then
               Retransmit RREQ
               IRet(i) = 1 {Update the node retransmission index IRet(i) by equating it to 1}
            End if
         End if
      End if
   End if
```

Figure 4. The proposed probabilistic algorithm.

## 7. SIMULATIONS AND RESULTS

In order to evaluate and analyze the performance of probabilistic flooding in noisy MANETs, four scenarios were simulated using the MANET simulator (MANSim) [12]. Before we proceed with the description of these scenarios, let us briefly introduce the simulation platform, MANSim. MANSim is a network simulator especially developed to evaluate and analyze the performance of a number of flooding optimization algorithms in MANETs. It is written in C++ programming language and it consists of four major modules: (1) Network module, (2) Mobility module, (3) Computational module, and (4) Algorithm module. In this work, we introduced some modification to the algorithm module to simulate probabilistic flooding in noiseless/noisy environment.

MANSim calculates a number of network performance measures, such as: the network reachability (RCH), the number of retransmission (RET), the average duplicate reception (ADR), average hop count (AHP), saved rebroadcast (SRB), and disconnectivity (DIS). These parameters are recommended by the Internet Engineering Task Force (IETF) group to judge the performance of the flooding optimization algorithms. Definition of these parameters can be found in [9]. However, in this work, we present results for two parameters only, these are: RCH and RET.

RCH is defined as the average number of reachable nodes by any node on the network normalized $n$. RCH can also be defined as the probability by which a RREQ packet delivered from source to destination node. RET is defined as the average number of retransmissions normalized to $n$. In addition, MANSim can be used to investigate the effect of a number of input parameters on the above performance measures, such as: $n$, $u$, $R$, $p_c$, $p_t$, etc.

### 7.1 Scenario#1: Investigate the Variation of RCH and RET against $p_c$ for Various fixed $p_t$'s

This scenario investigates the effect of $p_c$ on the performance (in terms of RCH and RET) of probabilistic algorithm of various fixed $p_t$'s (e.g., 0.7, 0.8, 0.9, and 1). $p_t$=1 means pure flooding. The input parameters for this scenario are given in Table 1. The simulation results for RCH and RET are plotted in Figures 5 and 6, respectively.

| Table 1. Input parameters. | | | | |
|---|---|---|---|---|
| Parameters | Scenario #1 | Scenario #2 | Scenario #3 | Scenario #4 |
| Geometrical model | Random distribution | Random distribution | Random distribution | Random distribution |
| Network area | 600x600 m | 600x600 m | 600x600 m | 600x600 m |
| Number of nodes ($n$) | 100 nodes | 75, 100, 125 nodes | 100 nodes | 100 nodes |
| Transmission radius ($R$) | 100 m | 75, 100, 125 m | 100 m | 75, 100, 125 m |
| Average node speed ($u$) | 5 m/sec | 5 m/sec | 2, 5, 8 m/sec | 5 m/sec |
| Size of mobility loop ($nIntv$) | 120 | 160, 120, 96 | 48, 120, 192 | 48, 120, 192 |
| Retransmission probability ($p_t$) | 0.7, 0.8, 0.9, 1.0 | 0.8 | 0.8 | 0.8 |
| Probability of reception ($p_c$) | 0.5-1.0 (Step 0.1) | 0.5-1.0 (Step 0.1) | 0.5-1.0 (Step 0.1) | 0.5-1.0 (Step 0.1) |
| Simulation time ($T_{sim}$) | 1800 sec | 1800 sec | 1800 sec | 1800 sec |
| Pause time ($\tau$) | $\tau = 0.75*(R/u)$ | $\tau = 0.75*(R/u)$ | $\tau = 0.75*(R/u)$ | $\tau = 0.75*(R/u)$ |

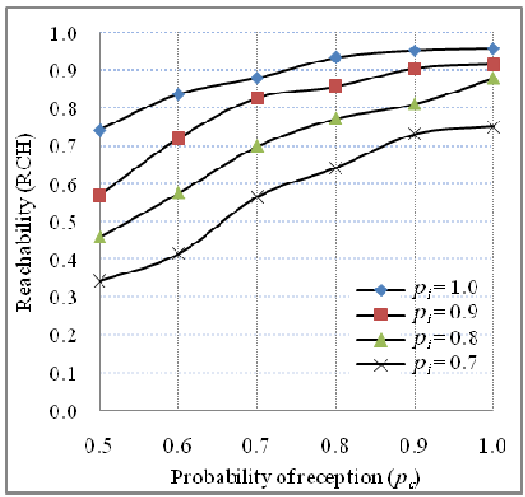
Figure 5. Variation of RCH against $p_c$ for various $p_t$.

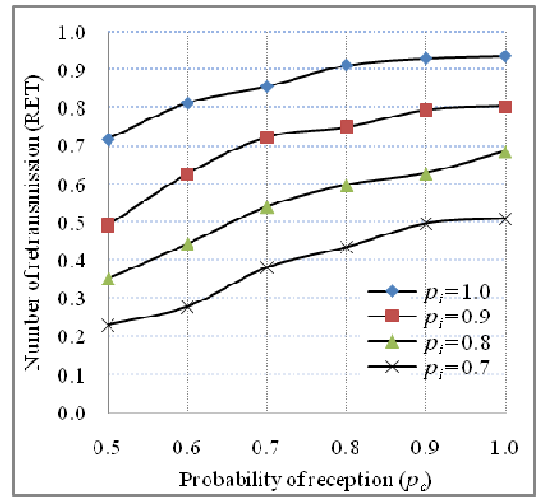
Figure 6. Variation of RET against $p_c$ for various $p_t$.

It can be seen from Figures 5 and 6 that RCH and RET are directly proportional to $p_c$. At the same time, for a certain $p_c$, both RCH and RET are decreased with decreasing $p_t$. This is very obvious from the discussion in Section 5.1. Although, the reduction in RET reduces the contention and also reduces the power and bandwidth consumption, which what we are looking for by using flooding optimization techniques; in this case, the reduction in RET cannot be considered as an advantage, because it comes along with significant reduction in the network RCH.

Let us define another parameter to deeply analyze the performance of the probabilistic algorithm and how does it understated due to the presence of noise, namely, the percentage relative change ($S_V$) of parameter $V$ (e.g., RCH or RET) at a certain $p_t$ can be calculated as:

$$S_V = \left( \frac{V(1) - V(p_c)}{V(1)} \right) \times 100 \qquad (1)$$

Where $V(1)$ and $V(p_c)$ are the computed parameter at $p_c=1$ (noiseless environment) and $p_c=p_c$ (noisy environment). Thus, it can be deduced from Figures 5 and 6 that $S_{RCH}$ and $S_{RET}$ increase as $p_t$ decreases, i.e., they are higher for small $p_t$ and lower for pure flooding. For example, when $p_c$ changes from 1 to 0.5, $S_{RCH}$ and $S_{RET}$ for $p_t=1$, 0.9, 0.8, and 0.7 are given in Table 2. The RCH of probabilistic algorithm suffers with increasing noise-level (decreasing $p_c$) and its suffering increases as $p_t$ decreases.

| Table 2. Scenario #1 The percentage relative change of $S_{RCH}$ and $S_{RET}$ |||||
|---|---|---|---|---|
| $p_t$ | $S_{RCH}$ (%) || $S_{RET}$ (%) ||
| | $p_c$=0.5 | $p_c$=0.6 | $p_c$=0.5 | $p_c$=0.6 |
| 1.0 | 22.5 | 12.5 | 23.5 | 13.2 |
| 0.9 | 37.9 | 21.4 | 38.9 | 22.0 |
| 0.8 | 47.9 | 34.6 | 48.8 | 35.5 |
| 0.7 | 54.5 | 44.9 | 54.9 | 45.5 |

### 7.2 Scenario#2: Investigate the Variation of RCH and RET against $p_c$ for various $n$

This scenario investigated the variation of RCH and RET against $p_c$ for various $n$. Three values of $n$ were investigated; these are: 75, 100, and 125 nodes. The $p_t$ of all nodes on the network is fixed at 0.8. All nodes were assumed to move with an average velocity of 5 m/sec. The input parameters for this scenario are given in Table 1. The variation of RCH and RET with $p_c$ for various $n$ are plotted in Figures 7 and 8. The results for $S_{RCH}$ and $S_{RET}$ are summarized in Table 3.

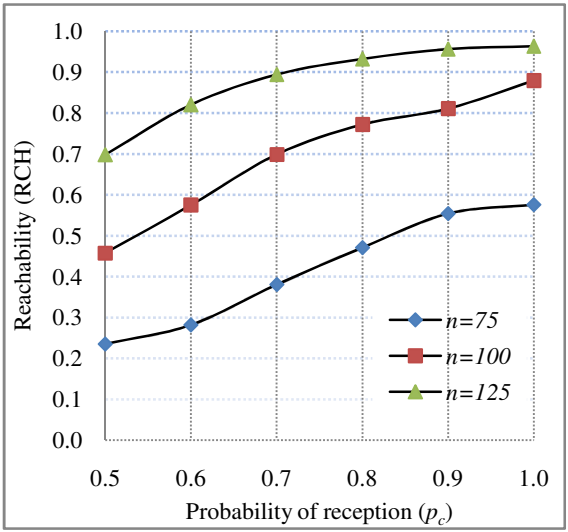
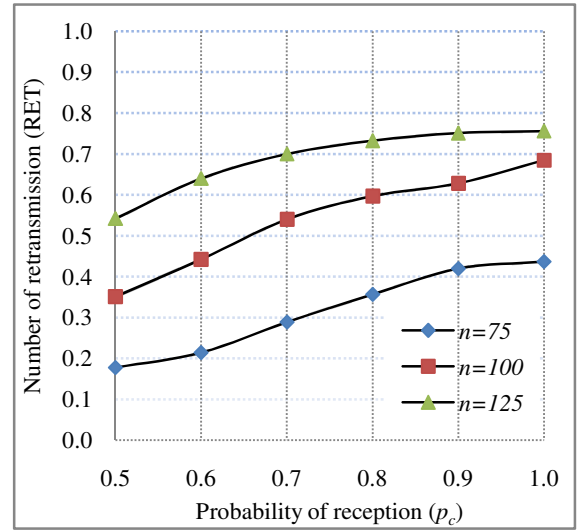

Figure 7. Variation of RCH against $p_c$ for various values of $n$.

Figure 8. Variation of RET against $p_c$ for various values of $n$.

| Table 3. Scenario #2 The percentage relative change of $S_{RCH}$ and $S_{RET}$ |||||
|---|---|---|---|---|
| $n$ | $S_{RCH}$ (%) || $S_{RET}$ (%) ||
| | $p_c$=0.5 | $p_c$=0.6 | $p_c$=0.5 | $p_c$=0.6 |
| 75 | 59.2 | 51.0 | 59.3 | 51.0 |
| 100 | 47.9 | 34.6 | 48.8 | 35.5 |
| 125 | 27.5 | 14.8 | 28.3 | 15.3 |

Before we precede with the explanation of the results in Figures 7 and 8, let us first defined $n_{avg}$ as a measure of nodes density, which is the average number of nodes that are covered by a transmitting node. It can be calculated as: $n_{avg}=\pi n R^2/A$, where $n$, $R$, $A$ are the number of nodes, radio transmission range of the node, and the network area. Thus, for this scenario, since $A$=600x600 and $n$=75, 100, 125, the values of $n_{avg}$ are 6.55, 8.73, and 10.91.

The results in the above two figures demonstrated that, for a network of $n$ nodes, RCH decreases as noise-level increases due to growing packet-loss, but the rate of change is less for dense networks, i.e., the performance of probabilistic algorithm is less affected by noise in dense networks. This can be explained as follows: The nodes covered by a transmitting node can act according to one of the following manners, either prohibits retransmission ($p_t$=0.8), or do not receive RREQ packet to retransmit due to packet-loss ($p_c$<1), or successfully receive and retransmit the RREQ packet to their neighboring nodes. As $n_{avg}$ increases, the probability of having successful nodes is higher, which means higher probability of successfully delivering the RREQ packet to the destination node contributing to higher RCH as shown in Figure 7. Some nodes fail to receive the RREQ packet, either due to packet-loss or their neighbors probabilistically prohibited from retransmission of RREQ packet, therefore, they have no packet to retransmit. This explains why RET decreases with increasing noise-level.

### 7.3 Scenario#3: Investigate the Variation of RCH and RET against $p_c$ for Various $u$

This scenario investigated the variation of RCH and RET against $p_c$ for various $u$. Three values of $u$ were investigated; these are: 2, 5, and 8 m/sec. The $p_t$ of all nodes on the network is fixed at 0.8. The input parameters for this scenario are given in Table 1. The variation of RCH and RET with $p_c$ for various $n$ are plotted in Figures 9 and 10.

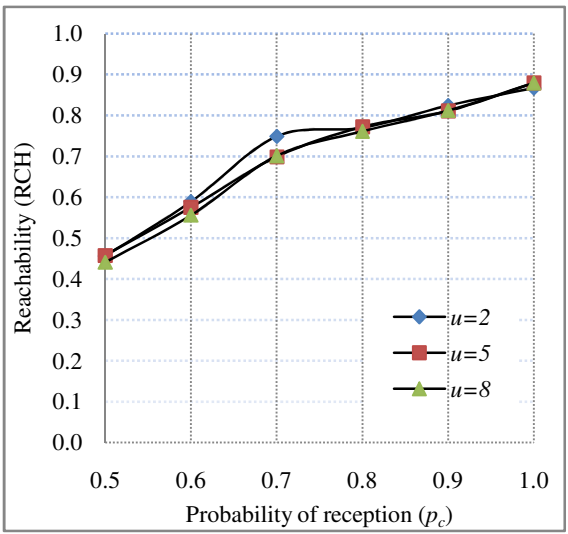
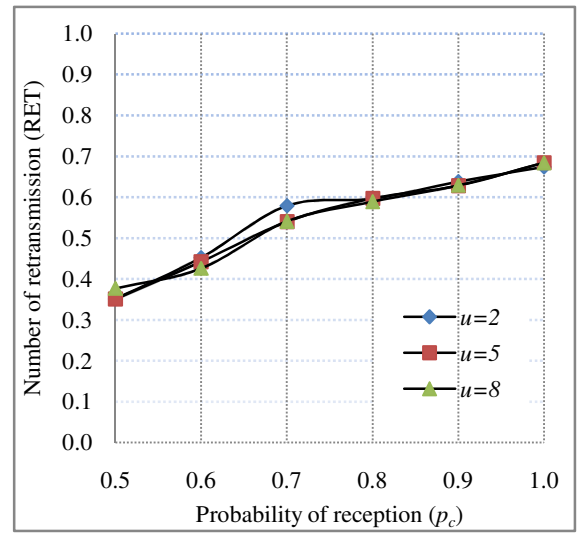

Figure 9. Variation of RCH against $p_c$ for various values of $u$.

Figure 10. Variation of RCH against $p_c$ for various values of $u$.

We can conclude from these two figures that RCH and RET decrease with increasing noise-level, which we have discussed above. The other conclusion is that the node velocity has almost no effect on RCH and RET. The reason for that can be explained as follows: suppose at time $t$, the node distribution is as shown in Figure 11a, where nodes A, B, C, and D, each has 8, 7, 9, and 4 first-hop neighbors. At time $t+\tau$ (Figure 11b), the node distribution is changed because all nodes may have randomly moved, and their first-hop neighbors changed to 7, 8, 4, and 9. This yields different actual behavior for the nodes A, B, C, and D, but most probably provides the same average network behavior and consequently RCH and RET.

### 7.4 Scenario#4: Investigate the Variation of RCH and RET against $p_c$ for Various $R$

This scenario investigated the variation of RCH and RET against $p_c$ for various $R$. Three values of $R$ were investigated; these are: 75, 100, and 125 m. The $p_t$ of all nodes on the network is fixed at 0.8. All nodes were assumed to move with an average velocity of 5 m/sec. The input parameters for this scenario are given in Table 1. The variation of RCH and RET with $p_c$ for various $R$ are plotted in Figures 12 and 13. For this scenario, the calculated values of $n_{avg}$ are 4.91, 8.73, and 13.64.

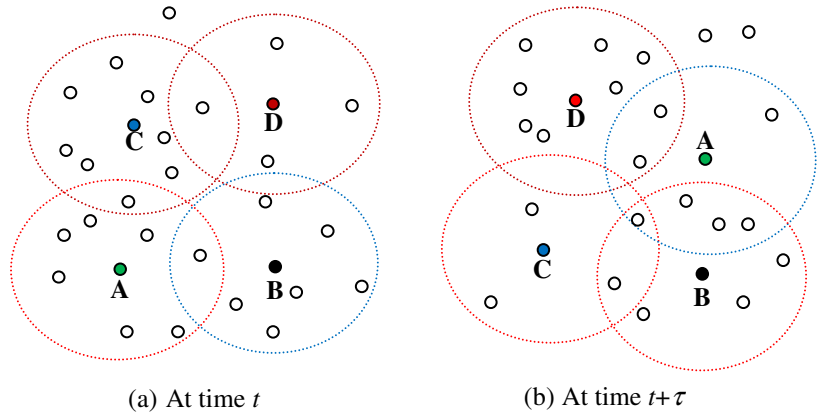

(a) At time $t$     (b) At time $t+\tau$

Figure 11. Node distribution at time $t$ and $t+\tau$.

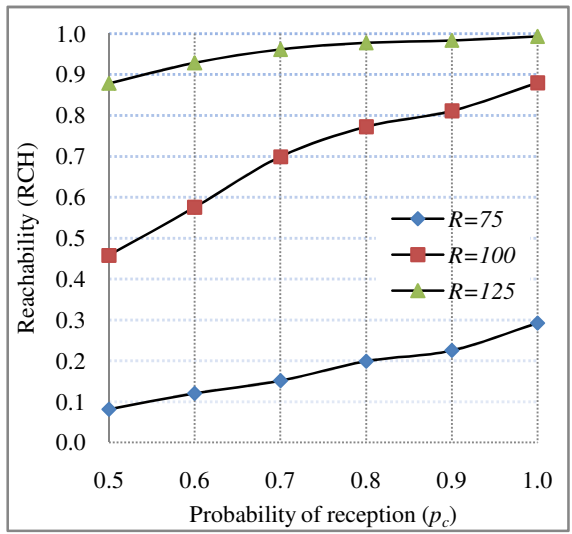

Figure 12. Variation of RCH against $p_c$ for various values of $n$.

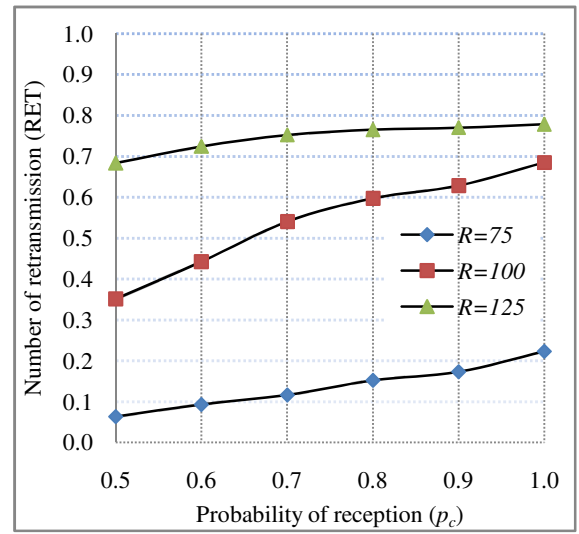

Figure 13. Variation of RET against $p_c$ for various values of $n$.

The results demonstrated that RCH decreases as noise-level increases due to growing packet-loss, but the rate of change is less for bigger $R$, i.e., the performance of probabilistic algorithm is better at bigger $R$. This is because as $R$ increases, the number of nodes that are covered by the transmitting node is higher. So that despite the fact that some of these nodes prohibit retransmission ($p_t$=0.8), others nodes do not receive RREQ packet to retransmit due to packet-loss ($p_c$<1), but the probability that some of them succeed in delivering the RREQ packet to their destination is higher and consequently enhancing RCH.

## 8. CONCLUSIONS

This paper presents a detail description of a simulation model that is developed to analyze the performance of a probabilistic algorithm for route discovery in noisy MANETs. The performance of the probabilistic approach against $p_c$ are simulated and analyzed for various $p_t$, $n$, $u$, and $R$. Based on the simulation results and discussions presented in Section 7, the main conclusions of this work can be summarized as follows: The network RCH always decreases with increasing noise-level. This is because of the high packet-loss introduced by increasing noise-level. Due to the high packet-loss, some of the nodes fail to receive the RREQ packet, therefore, they have no packet to retransmit and consequently RET is decreased.

Scenario #1 showed that pure flooding or high $p_t$ probabilistic algorithm are less affected by the presence of noise, as $S_{RCH}$ is less for pure flooding and it decreases with decreasing $p_t$. Scenario #2 and #4 demonstrated that increasing $n$ and/or $R$ reduces the effect of noise on the performance (RCH) of the probabilistic algorithm ($S_{RCH}$ is less for higher $n$ and/or $R$), while as shown in scenario #3, the node mobility has no or insignificant effect on the performance of the algorithm.

## AUTHORS

***Hussein Al-Bahadili*** (*Hbahadili@aabfs.org*) received his B.Sc degree in Engineering from University of Baghdad, Iraq, in 1986. He received the M.Sc and PhD degree from University of London (Queen Mary College), UK, in 1988 and 1991, respectively. He is currently an associate professor at the Arab Academy for Banking & Financial Sciences (AABFS), Jordan. He is also a visiting researcher at the Centre of Wireless Networks and Communications (WNCC), University of Brunel (UK). He has published many papers in leading journals and world-level scholarly conferences. He recently published two chapters in books in IT. His research interests include computer networks design and architecture, routing protocols optimizations, parallel and distributed computing, cryptography and network security, and data compression.

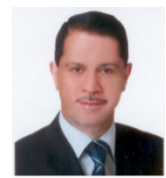

***Khalid Kaabneh*** (*kkaabneh@aabfs.org*) received his Bachelor's degree in Electrical Engineering and Computer Engineering in 1989. In 1991, he earned an MS Degree in Management Information Systems. A PhD. was bestowed upon him in 2001 for his research in the area of Multimedia Audio Watermarking. All degrees were awarded from The George Washington University in Washington DC, USA. He is currently working as an associate professor at the Arab Academy for Banking & Financial Sciences, Jordan. His research interests include areas like digital communications, wireless networks, digital multimedia, copyright protection, and data and system security. Dr. Kaabneh is an approved computer security consultant and published various security research papers in a multitude of local and international journals.

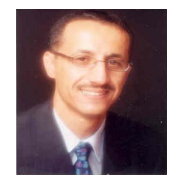